\documentclass[twocolumn,showpacs,amsmath,amssymb,pre,floatfix]{revtex4-1}

\pdfoutput=1

\usepackage{graphicx}
\usepackage{bm}
\usepackage{bbold}
\usepackage{amsmath,amsfonts,amssymb}

\begin{document}

\title{Signatures of many-body localisation in a system without disorder and the relation to a glass transition}

\author{James M.\ Hickey}
\author{Sam Genway}
\author{Juan P.\ Garrahan}
\affiliation{School of Physics and Astronomy, University of
Nottingham, Nottingham, NG7 2RD, UK}


\date{\today}

\begin{abstract}
We study a quantum spin system---adapted from a facilitated spin model for classical glasses---with local bilinear interactions and without quenched disorder which seems to display characteristic signatures of a many-body localisation (MBL) transition.  From direct diagonalisation of small systems, we find a change in certain dynamical and spectral properties at a critical value of a coupling, from those characteristic of a thermalising phase to those characteristic of a MBL phase.  The system we consider is known to have a quantum phase transition in its ground-state in the limit of large size, related to a first-order active-to-inactive phase transition in the stochastic trajectories of an associated classical model of glasses.  Our results here suggest that this first-order transition in the low-lying spectrum may influence the 
rest of the spectrum of the system in the large size limit.  These findings may help understand the connection between MBL and structural glass transitions.    
\end{abstract}

\maketitle

\section{Introduction}

For over half a century, it has been understood that a single quantum particle can become localised in space in the presence of a disordered potential~\cite{Anderson1958, Lee1985}.  Recently, there has been a surge of interest in localisation in the context of interacting many-body systems~\cite{Basko2006,*Basko2007,Oganesyan2007,*Pal2010,Znidaric2008, *Monthus2010,*Vosk2013,*Lev2014,*Pollmann2014,*Pollmann2012,*Nayak2013,[For a recent review see ]Nandkishore2014}.  When a closed interacting system exhibits many-body localisation (MBL), a breakdown of thermalisation \cite{Polkovnikov2011} occurs: the system is unable to function as its own thermal bath, dynamics retains memory of its initial state, and expectation values of observables do not relax to the values expected from thermal equilibrium \cite{Nandkishore2014}.  To date, MBL has been demonstrated mostly in systems with quenched disorder, and it is currently of interest~\cite{Nandkishore2014} to establish whether MBL is exhibited in quantum systems where the Hamiltonian itself is translationally invariant 
\cite{Nandkishore2014,Yao,Horssen2015,Schiulaz2015,Papic2015}.  MBL may be possible in systems with both fast- and slow-moving particles~\cite{Giuseppe2012,*Grover2013,*De-Roeck2013,*Schiulaz2013,*Huveneers2014,*Muller2014}, where the slow particles provide an effective disordered potential in which the fast particles can appear localised.  Whether MBL can be seen without such impurities providing effective disorder is an open question~\cite{Nandkishore2014}.

Here we present and study a quantum spin system which has local interactions and is free of disorder that appears to show the characteristic features of MBL.  The Hamiltonian we consider is related to a deformation of the master operator of a classical glass model, specifically the one-spin facilitated Fredrickson-Andersen model \cite{Fredrickson1984,Ritort2003}.  This model is  
known \cite{Garrahan2007,*Garrahan2009} to display a (first-order) phase transition in its largest eigenvalue.
In the classical stochastic context this corresponds to a singularity in the cumulant generating function of the dynamical activity \cite{Lecomte2007,Baiesi2009} and thus an indication of a non-equilibrium transition to an inactive non-equilibrium glass state \cite{Garrahan2007,Hedges2009,[For a review of this perspective on the glass transition problem see ]Chandler2010}.  In the quantum context this singularity corresponds to a quantum phase-transition in the ground-state of the system.  We show below that this singular structure in the low-lying part of the spectrum of the Hamiltonian seems to have also an influence on the rest of the spectrum, giving rise to unitary dynamics that display some of the hallmarks of a MBL transition.  
Our results provide a connection between mechanisms for slow relaxation in models of classical glasses, and mechanisms for slow thermalisation and non-ergodicity in closed quantum systems under unitary evolution.

\section{Model}

The system we study consists of spins or qubits on the sites of a one-dimensional lattice with periodic boundary conditions.  The Hamiltonian has the form,
\begin{equation}
H = - \frac{1}{2} \sum_{i=1}^{N} 
\left( \Gamma \sigma_{i}^{x} - \gamma \sigma_{i}^{z} - \kappa \right)
\left(\sigma_{i-1}^{z} + \sigma_{i+1}^{z} + 2 \lambda \right) ,
\label{H}
\end{equation}
where $i=1,\ldots,N$ are the sites of the lattice, $\sigma_{i}^{x,y,z}$ are the usual Pauli operators acting on site $i$, and the parameters $(\Gamma,\gamma,\kappa,\lambda)$, which quantify the strength of the various fields and couplings, are uniform throughout the lattice (i.e., there is {\em no quenched disorder}). 

We write the Hamiltonian as in (\ref{H}) to highlight the connection to so-called kinetically constrained models of glasses \cite{[For reviews see: ]Ritort2003}. When $\kappa+\gamma=\Gamma^{2}$ and $\lambda=1$ the operator $H$ is the symmetrized version 
of the master operator of a Fredrickson-Andersen (FA) facilitated spin model \cite{Fredrickson1984,Ritort2003,Jack2006} (see below). 
In this case, the first factor in each term of (\ref{H}) is the local operator that flips a spin at site $i$.  The second factor constrains the rate at site $i$ to the state of the spins at sites $i \pm 1$, so that changes at site $i$ cannot occur if both these spins are in the down state (in the $z$ basis). 
For $\lambda > 1$ this constraint is softened, and we have the so-called soft-FA model, see \cite{Elmatad2010}.  When $\gamma+\kappa \neq \Gamma^{2}$, the operator $H$ is related to a deformation of the master operator for the dynamics of the FA (or soft-FA) model, which is known to show a singular change in its ground state at some value of $\Gamma$ \cite{Garrahan2007,Elmatad2010}, as we discuss below.

\begin{figure}
\includegraphics[width=1.0\columnwidth]{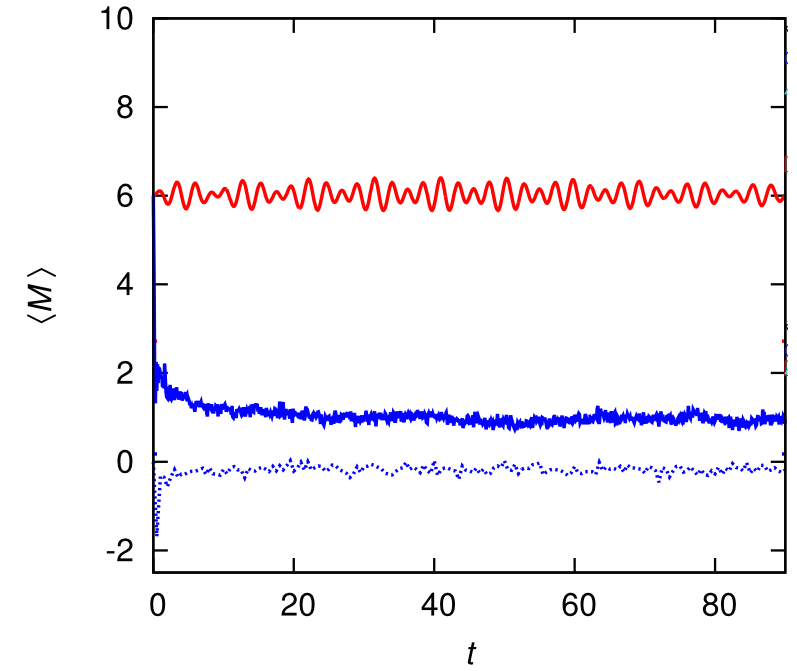}
\caption{
{\em Relaxation of an observable.} 
Time evolution of the average magnetisation, $\langle M \rangle_t$. 
The system has size $N=16$, with $\epsilon=0.7$ and $\lambda=1$.  
We show $\langle M \rangle_t$ from an initial state with well defined magnetisation $M_{0}$ at different values of $s$ either side of $s_*=0$: $s=-2$ with $M_{0}=6$ (full blue curve), $s=2$ with $M_{0}=6$ (full red curve), and $s=-2$ with $M_{0}=0$ (dotted blue curve). For $s<0$ the magnetisation seems to relax quickly towards similar values despite the difference in initial state, while for $s>0$ it appears not to relax at all.}
\label{fig1}
\end{figure}

We parametrize the couplings in the following way, 
\begin{equation}
\Gamma = e^{-s} \sqrt{\epsilon}, \;\;
\gamma = \frac{1}{2} (1-\epsilon), \;\;
\kappa = \frac{1}{2} (1+\epsilon) .
\label{e}
\end{equation}
When $s=0$, $H$ is equivalent to a classical stochastic operator.  
In the quantum language we would say that in this case we are at a Rokhsar-Kivelson point \cite{Rokhsar1988,Castelnovo2005}: 
the ground-state, whose energy vanishes, is given by the ``square root'' of the equilibrium probability of the stochastic process, 
in this case the direct product, 
\[
| \sqrt{{\rm eq.}} \rangle \equiv (1+\epsilon)^{-N/2} \bigotimes_{i} ( | 0_{i} \rangle + \sqrt{\epsilon} | 1_{i} \rangle ) ,
\]
 where $| n_{i}=0,1 \rangle$ are the eigenstates of $\sigma_{i}^{z}$ with eigenvalues $-1$ and $+1$, respectively, with $\epsilon$ the relative weight between a spin being up to being down in this basis.

We now expand on the connection to a stochastic problem when $s=0$.  The master operator that generates the stochastic dynamics of the FA model is \cite{Jack2006,Elmatad2010}
\begin{eqnarray}
W = \sum_{i} 
\left[ \epsilon \sigma_{i}^{+} + \sigma_i^- - \epsilon \bar{n}_i - n_i \right] 
\nonumber \\
\times 
\left( n_{i-1} + n_{i+1} + \lambda - 1 \right) , 
\label{W}
\end{eqnarray}
where $\sigma^z_{i} = 2 n_{i} -1$ and $\bar{n}_{i} = 1 - n_{i}$.  
In the classical problem the rate $\epsilon$ is typically determined by temperature, e.g., $\epsilon = e^{-1/T}$.
A stochastic operator such as $W$ is in general non-Hermitian.  The Perron-Frobenius theorem implies that the largest eigenvalue of $W$ is zero, and all the other eigenvalues are real and negative (the negative of the rates of relaxation).  The zero eigenvalue is a consequence of probability conservation.  The corresponding right eigenvector of $W$ is the stationary states, which for the generator above is   
the equilibrium probability
\[
| {\rm eq.} \rangle \equiv \left(1+\epsilon \right)^{-N} \bigotimes_{i} ( | 0_{i} \rangle + \epsilon | 1_{i} \rangle ) .
\]
The associated right eigenstate is the ``flat'' or ``trace'' state
\[
\langle - | \equiv \bigotimes_{i} ( \langle 0_{i} | +  \langle 1_{i} | ) ,
\]
which when applied to a probability vector simply gives the sum over all configurations.  Notice that the normalisation of $| {\rm eq.} \rangle$ above implies
$\langle - | {\rm eq.} \rangle = 1$.

The connection between $W$ and $H$ is via a similarity transformation. From the equilibrium probability vector one can construct the diagonal operator 
\[
P = \bigotimes_{i} ( | 0_{i} \rangle \langle 0_{i} | + \epsilon^{-1/2} | 1_{i} \rangle \langle 1_{i} | ) .
\]
The Hermitian  $H$ is obtained from the stochastic $W$ via a the transformation, $H = - P^{-1} W P$.  This implies that $H$ and $W$ have the same spectrum.  Notice also that (up to normalisation) $| {\rm eq.} \rangle = P | \sqrt{{\rm eq.}} \rangle$
and $\langle - | = \langle \sqrt{{\rm eq.}} | P^{-1}$. 

The operator $W$ can be ``deformed'' or ``tilted'' in order to extract the statistics of time-integrated observables of the dynamics generated by $W$ \cite{Garrahan2007}. For example, if the observable of interest is the number of spin flips in a trajectory (the dynamical activity \cite{Lecomte2007,Baiesi2009}), then the corresponding tilted generator is 
\begin{eqnarray}
W_{s} = \sum_{i} 
\left[ e^{{-s}} \left( \epsilon \sigma_{i}^{+} + \sigma_i^- \right) - \epsilon \bar{n}_i - n_i \right] 
\nonumber \\
\times 
\left( n_{i-1} + n_{i+1} + \lambda - 1 \right) .
\end{eqnarray}
While this tilted operator is also non-Hermitian, it can be made Hermitian by the same similarity transformation as above to give the, giving the general Hamitonian of Eq.\ (\ref{H}) for arbitrary $s$.

When $s \neq 0$ we are away from the RK point and the ground-state energy of $H$ may no longer vanish.  This is equivalent to the statement that the operator $W(s)$ is no longer a stochastic operator for $s \neq 0$.  As $s$ is increased, and depending on the value of $\epsilon$, there can be a change in the ground-state.  In the large size limit, $N \to \infty$, this change may become singular at some $s_*(\lambda)$.  For $\lambda \to 1$ this occurs for all $\epsilon$ at $s_* \to 0$, where the change is from the ``equilibrium'' state $| \sqrt{{\rm eq.}} \rangle$, which dominates at $s = 0^{-}$, to the ``inactive state'' $| {\rm in.} \rangle \approx \bigotimes_{i} | 0_{i} \rangle$, which dominates at $s=0^{+}$ 
\footnote{When {$\lambda=1$} the state {$\bigotimes_{i} | 0_{i} \rangle$} is an eigenstate with zero eigenvalue for all $s$ and forms a one-dimensional subspace of $H$.  Thus we remove it from the spectrum; the quantum phase transition occurs in the ``connected component'' of $H$: the support of {$| {\rm in.} \rangle$} is on states {$| \mathbf{n} \rangle$} with a subextensive number of $n_{i}=1$. In the large system size limit the difference between {$| {\rm in.} \rangle$} and {$\bigotimes_{i} | 0_{i} \rangle$} becomes vanishingly small~\cite{Garrahan2007,Elmatad2010}.  For $\lambda > 1$, {$\bigotimes_{i} | 0_{i} \rangle$} ceases to be a trivial eigenstate, and all {$| \mathbf{n} \rangle$} are connected by $H$.
}.
This quantum phase transition in $H$ is first-order \cite{Garrahan2007,Elmatad2010}.  

The ground-state transition from $| \sqrt{{\rm eq.}} \rangle$ to $| {\rm in.} \rangle$ at $s_*=0$ (we consider $\lambda=1$ from now on for simplicity) has the flavour of a localisation transition in the Fock basis $| \mathbf{n} \rangle \equiv \bigotimes_{i} | n_{i} \rangle$, since $| \sqrt{{\rm eq.}} \rangle$ is spread as a probability over the states $| \mathbf{n} \rangle$ while $| {\rm in.} \rangle$ is highly concentrated on $| 0 \cdots 0 \rangle$.  In the classical context this is a non-equilibrium transition from a relaxing ``liquid'' to a non-ergodic ``glass'' \cite{Garrahan2007,Hedges2009}.

\begin{figure}
\includegraphics[width=1.0\columnwidth]{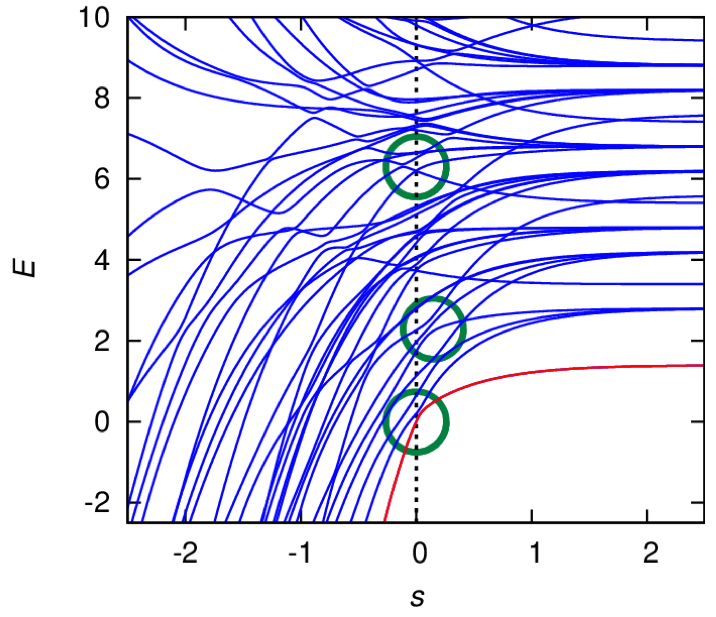}
\caption{{\em Change in the spectrum.}
Spectrum of $H$ for a system of size $N=9$, with $\epsilon=0.7$ and for $\lambda=1$. We only show eigenvalues with zero momentum (see text).  The ground state, indicated in red, is known to display a first-order singularity in the large size limit.  Circles indicate avoided crossings occurring near $s_*=0$ which may become singular as well as $N \to \infty$.}
\label{fig2}
\end{figure}

\section{Signatures of MBL}

We now study signatures of a possible MBL transition in our system.  Due to translational invariance, the definition of a MBL transition in terms of entanglement entropy scaling, as in Ref.~\cite{Nayak2013}, or through the decay of connected correlators, is not applicable.  Instead we investigate the characteristics of the many-body eigenstates and compare their properties with those associated with both MBL and thermal states.  We first consider the thermalisation properties by examining the time evolution of global observables under the unitary dynamics generated by $H$. Figure 1 
shows the average total magnetisation in the $z$-direction, $\langle M \rangle_t$, as a function of time, where 
\[
M \equiv \sum_{i} \sigma_{i}^{z},
\]
and $\langle \cdot \rangle_t$ indicates expectation value in the state 
\[
|\psi(t)\rangle = e^{-i t H} | {\rm init.} \rangle 
\]
(where $\hbar=1$).
In Fig.\ 1 the initial state $| {\rm init.} \rangle$ for all 
curves is a zero momentum state with well defined total magnetisation $M_{0}$.  There are two clearly distinct regimes  depending on the value $s$.  For $s<0$, $\langle M \rangle_t$ relaxes on a relatively short time scale (much shorter that the renewal timescale for the finite $N$ systems we simulate), as shown for two initial states, $M_{0}=6$ (full blue curve) and $M_{0}=0$ (dotted blue curve).  These initial states were chosen so that their average energies $\langle {\rm init.} | H | {\rm init.} \rangle$ were as close as possible in this finite sized system.  The value to which $\langle M \rangle_t$ tends in the long time is also close (within fluctuations associated with a finite system), despite the fact that $M_{0}$ is very different.  One would associate this behaviour with conditions where observables thermalise \cite{Polkovnikov2011} to a level dictated by their  energy.  In contrast, for $s>0$, the magnetisation remains close to the value in the initial state for long times, a behaviour associated with absence of thermalisation.
This change takes place at $s_*=0$, the value of $s$ of the transition in the ground state of $H$. Note that the $| {\rm init.} \rangle$ chosen are ``atypical'' initial states \cite{Nandkishore2014}.  Similar behaviour is observed for other atypical initial conditions. 

The result of Fig.\ 1 is a first indication that the ground-state transition at
$s_*=0$ may actually affect the bulk of the spectrum.  To illustrate
this we show in Fig.\ 2 the spectrum of $H$ (for a small system for clarity), as a function of $s$ for fixed $\epsilon$ and $\lambda=1$. 
The ground state is indicated by red.  We show only the zero momentum sector to
avoid trivial eigenvalue crossings due to the fact that the system is
translationally invariant.  We have also removed the isolated state
{$\bigotimes_{i} | 0_{i} \rangle$} \cite{Note1}.  [Some crossings remain in
Fig.\ 2 due to residual discrete symmetries which are more difficult to
remove from the spectrum.]  Around $s=0$ there is a crossover, associated with
the avoided crossing indicated by a circle.  When $N \to \infty$ this crossover becomes the quantum phase transition of $H$ at $s_*=0$.  From Fig.\ 2 we see that near $s=0$ there seems to be a proliferation of other avoided crossings (indicated by circles) in  other eigenstates.  This may be characterized by examining the spectral statistics of the model, as discussed below in Sec.\ IV.

\begin{figure}
\includegraphics[width=1.0\columnwidth]{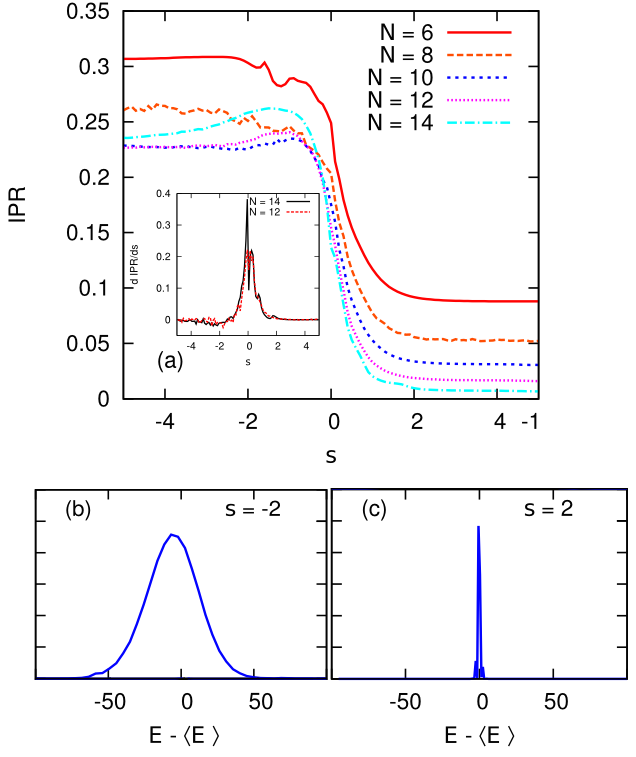}
\caption{
{\em Localisation properties of eigenstates in Fock basis.} (a) Average inverse participation
ratio as a function of $s$ for various system sizes.   (b,c) Average local
density of states at two values of $s$ at either side of the MBL transition: for $s<0$ the LDOS is approximately Gaussian, while for $s>0$ is highly concentrated.
}
\label{fig3}
\end{figure}

Examining the spectrum in Fig.\ 2, we find in the potentially non-thermalising phase ($s > 0$) the energy eigenstates form clusters, 
where every eigenstate within each cluster has a well defined integer magnetisation $\langle M \rangle$. 
This energy level structure has been observed in other MBL systems~\cite{Giuseppe2012,*Grover2013,*De-Roeck2013,*Schiulaz2013,*Huveneers2014,*Muller2014}, and potentially indicates that in the non-thermalising regime the energy eigenstates may possess characteristics of the many-body Fock basis, i.e.\ $\langle M \rangle \in \mathbb{Z}$.  Furthermore, preparing a system at fixed energy, the long time dynamics (dominated by the intra-cluster energy eigenvalues) only has contributions from a small subset of Fock states with similar $\langle M \rangle$. 

Our final characterization of a possible MBL transition is by considering how the eigenstates are distributed over the Fock basis states $| \mathbf{n} \rangle$.  We quantify this as usual via the inverse participation ratio (IPR), which for an eigenstate $| E \rangle$ reads, 
\[
{I}(E) \equiv \frac{1}{2^{N} \sum_{\mathbf{n}} | \langle E | \mathbf{n} \rangle |^{4}} 
\]
which implies that ${I}(E)$ is $O(2^{-N})$ if $| E \rangle$ is highly concentrated on some $| \mathbf{n} \rangle$, and $O(1)$ otherwise. Figure 3(a) shows the average of the IPR over the bulk of the spectrum, 
\[
\bar{I} \equiv 2^{-N} \sum_{E} {I}(E), 
\]
as a function of $s$
\footnote{The IPR is averaged over a Gaussian window centred on the central eigenenergies and of width $\sim 100\Delta$, where $\Delta$ is the mean energy level spacing.}.  There is a clear crossover in the behaviour of $\bar{I}$ near $s_*=0$.  For $s<0$ the average IPR is of $O(1)$, indicative of eigenstates spread out in $| \mathbf{n} \rangle$.  For $s>0$, $\bar{I}$ becomes small, seemingly tending to zero with $N$, indicative of eigenstates localised in $| \mathbf{n} \rangle$.  The change also appears to sharpen with increasing $N$, consistent with a discontinuity at $s_*=0$ in the large size limit.  This apparent sharpening is corroborated with a peak in the derivative of $\bar{I}$ at $s = 0$ which grows with increasing system size, see the inset of Fig. 3(a).

In Figs.\ 3(b,c) we show the average local density of states (LDOS).  
This is defined from the local density of states, 
\[
\text{LDOS}(E|\mathbf{n}) \equiv \sum_{E'} |\langle E'|\mathbf{n}\rangle|^2\delta(E-E'),
\]
by centering each $\text{LDOS}(E|\mathbf{n})$ around its average $E$ and then averaging over $\mathbf{n}$ \footnote{Note that $\text{LDOS}(E|\mathbf{n})$ is a function of $\mathbf{n}$'s energy, thus by centering the distribution first and then averaging we obtain an average LDOS, cf.~\cite{Cohen2003}.}. 
For $s<0$ the distribution over eigenstates appears Gaussian (indicative of a thermalising phase~\cite{Genway2013,*Genway2012}), while for $s>0$ it is highly concentrated on an eigenstate.  This indicates that in the non-thermalising phase the eigenstates are localised over a few many-body Fock states with that energy.  The crossover from a distribution which is close to a delta function to one which is Gaussian is similar to the change one would encounter in the  LDOS of an integrable quantum system (where thermalisation is not expected to occur) when a banded non-integrable perturbation is added (such that the perturbed system would thermalise)~\cite{Genway2013,*Flambaum2000,*Flambaum2001}.  In such a system one would expect that the average IPR $\bar{I}$ is approximately $\frac{1}{3 \sqrt{2}}~(\sim 0.235)$ (note that this is less than the $\bar{I}$ which one would expect for random eigenstates described by the Porter-Thomas distribution).  The extracted $\bar{I}$ is close to this value, highlighting the connection between the LDOS, the average IPR and non-integrable perturbations in integrable systems.

\begin{figure}
\includegraphics[width=1.0\columnwidth]{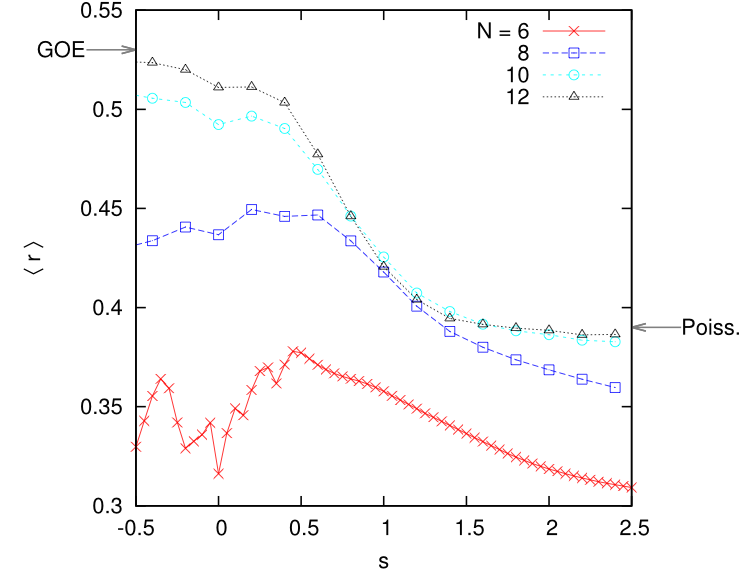}
\caption{
{\em Level spacing statistics with added disorder.} 
Average gap between adjacent energy levels, $\langle r \rangle$, as a function of $s$ (averaged over the whole spectrum), for the a disordered version of Eqs.\ (\ref{H},\ref{e}) according to Eq.\ (\ref{ei}).  For increasing $N$ there is an increasingly sharp crossover from a value of $\langle r \rangle$ compatible with GOE statistics to one compatible with Poisson statistics, as is usually found in disordered systems displaying thermal-MBL transitions. The crossovers seems to be around $s=1$ in the disordered case. 
In the figure $\epsilon = 0.7$ and $g=0.1$.  
}
\label{fig4}
\end{figure}

\section{Spectral Statistics and Disorder}

We find that the spectral statistics of the zero momentum eigenstates change in character at $s=0$.  This change is not the usual one between GOE and Poissonian, as this sector contains extra symmetries which are difficult to identify.  However, 
we can remove this symmetries by adding a small amount of disorder.  We can do so, as described below, in a way that is known not to change significantly the corresponding stochastic dynamics of the classical problem, in the hope that this change does not alter qualitatively the behaviour of the quantum problem.  

Lets consider first the disorderless $H$ of Eq.\ (\ref{H}). 
To quantitatively study the apparent emergence of avoided crossings in the spectrum as we cross $s = 0$ we examine the level spacing statistics.  The level spacings are defined as $\delta^{(n)} = |E^{(n)}-E^{(n-1)}|$ where $E^{(n)}$ is the many-body eigenenergy of the eigenstate $n$ and an important quantity of interest is the ratio of adjacent gaps 
\[
r^{(n)} = \frac{\min\{\delta^{(n)},\delta^{(n+1)}\}}{\max\{\delta^{(n)},\delta^{(n+1)}\}} .
\]
If the spectrum possesses Gaussian-orthogonal ensemble (GOE) statistics the average of this quantity will converge to $\langle r \rangle \rightarrow 0.53$ in the large system size limit~\cite{Pal2010}.  Similarly if the level statistics are Poissonian it will instead converge to $\langle r \rangle \rightarrow 0.39$. 
This means that by examining this average as a function of $s$ we can see if the avoided crossings do emerge as $s$ changes from positive to negative values. 

Applying this to $H$ of Eq.\ (\ref{H}), specifically examining the $k = 0$ momentum sector, we find that there is indeed a crossover in $\langle r \rangle$ at $s = 0$, suggesting a change in the level spacing statistics.   However, for $s > 0$ we find $\langle r \rangle \lesssim 0.3$ indicating that spectral statistics are ``more than Poissonian'', as expected from the presence of extra conserved quantities are in the system, at least in the $k = 0$ momentum sector which we explore.  Tuning $s$ towards negative values, the average level spacing ratio increases, suggesting that the spectrum is becoming less Poissonian than in the putative non-ergodic phase of $s>0$: avoided crossings emerge but it is not exactly GOE.  These extra symmetries are difficult to identify and remove in the translationally invariant case of Eq.\ (\ref{H}).  

In order to explore the level statistics more carefully we consider adding a small amount of disorder which we argue should not change the physics of the problem in any substantial way.  Consider the case where the rate constants $\gamma$ and $\kappa$ of Eq.\ (\ref{H}) become site dependent through the coupling $\epsilon$, see Eqs.\ (\ref{e}), acquiring a small amount of (quenched) disorder:
\begin{equation}
\epsilon \rightarrow \epsilon_i = |\epsilon - g~\pi_i| ,
\label{ei}
\end{equation}
where $i$ is the site index, $g$ is the strength of the disorder and $\pi_i$ is a Gaussian distributed random variable of mean $0$ and variance $1$, which is independent from site to site.  From the point of view of the classical dynamics generated by the operator $W$ at $s=0$, the introduction of this disorder is not very significant: the classical dynamics still tends to a stationary state which is a product state, but where there is a slight site to site variation of the average excitation concentration,
\[
| {\rm eq.} \rangle \rightarrow  \bigotimes_{i} 
\left(1+\epsilon_{i} \right)^{-1}
( | 0_{i} \rangle + \epsilon_{i} | 1_{i} \rangle ) .
\]
Of course, the classical stochastic dynamics is dominated by the leading eigenstates of $W$, while the quantum dynamics generated by $H$ depends on the whole spectrum.  Nevertheless, we expect that a small amount of disorder should not affect too much the properties of the quantum dynamics.   

In the presence of this quenched disorder translation invariance is broken, and we can expect, if there is a thermal-MBL change in the system, to see a corresponding GOE-Poisson change in the spectral statistics.  Diagonalising the Hamiltonian, for $\epsilon = 0.7$ and $g = 0.1$, we find the spectrum is comparable to the translationally invariant case, however the transition point appears to shift closer to $s = 1$.  Averaging the ratio of adjacent gaps over $100$ realisations of the disorder we find a crossover in the spectral statistics from values compatible with GOE ($s < 1$) to Poissonian ($s > 1$), see Fig.\ 4.  The shift in the apparent transition from around $s=0$ in the clean case to around $s=1$ in the disorder case suggests that the presence of disorder has a stronger effect in the excited states than in the low-lying spectrum, the latter being relevant for the unitary dynamics while only the former is relevant for the classical stochastic dynamics.

\bigskip

\section{Discussion}

The evidence above suggests that in the quantum system described by the
Hamiltonian (\ref{H}), in the case where $\lambda = 1$, there is a MBL
transition occurring at $s_*=0$ for all values of the parameter $\epsilon$. 
Quantities such as the IPR and the LDOS, 
give indications that there might be a crossover
from an ergodic thermalising phase to a non-ergodic non-thermalising phase as we change $s$. 
In what appears to be the non-ergodic phase the eigenstates have contributions from a finite number of Fock states and so are
localized on the Fock basis.  Our finite-size results also suggest that this transition sharpens to become a first-order transition in the thermodynamic limit, but of course this is hard to establish for the limit range of sizes accessible to our simulations.  Since $H$ is related to the deformed (or tilted \cite{Touchette2009}) generator for the dynamics of the classical FA glass model \cite{Garrahan2007}, our results here connect the (first-order) non-equilibrium glass transition in that model to a potential MBL transition in the quantum problem.  Here we have studied a quantum Hamiltonian related to the classical facilitated FA glass model.  Analogous signatures of a thermal-MBL crossover/transition are found in a quantum version of the East model, see Ref.\ \cite{Horssen2015}.

When $\lambda>1$, the Hamiltonian (\ref{H}) is associated to the soft-FA model \cite{Elmatad2010}.  
From its dynamics we know that in this case $H$ may or may not have a ground
state transition depending on $\lambda$ and $\epsilon$.  For $\lambda - 1
\gtrsim 0$ the ground state transition is present \cite{Elmatad2010}, and we would expect a MBL transition
 across the spectrum as in the $\lambda=1$ case, but occurring at some $s_{*}(\lambda,\epsilon) \neq 0$.  
 In contrast, when $\lambda - 1$ is large enough the ground state transition disappears \cite{Elmatad2010}, 
 allowing for an interesting situation.  In the quantum version of this problem this may be connected to an inverted ``mobility edge'' \cite{Nandkishore2014}.

Other kinetically constrained glass models have active-inactive transitions \cite{Garrahan2007,Elmatad2010}.  One such class of systems which would be interesting to consider in the quantum context are constrained lattice gases \cite{Ritort2003}, where hopping between sites is constrained by the state of neighbouring sites.  If their associated quantum problem also displays a MBL transition, like the one we showed here for systems based on facilitated spin models, then this MBL transition would be one where also particle transport ceases in the localised phase.

\bigskip

\acknowledgments
This work was supported by Leverhulme Trust grant No.\ F/00114/BG and by EPSRC Grant no.\ EP/M014266/1.

\bibliography{mabolo}

\end{document}